# Sampling Online Social Networks: Metropolis Hastings Random Walk and Random Walk

Xiao Qi[1*]

[1] School of Social Science, the University of Southampton, University Road, Southampton, SO17 1BJ, United Kingdom.
*Corresponding author xq4u19@soton.ac.uk

## Abstract

As social network analysis (SNA) has drawn much attention in recent years, one bottleneck of SNA is these network data are too massive to handle. Furthermore, some network data are not accessible due to privacy problems. Therefore, we have to develop sampling methods to draw representative sample graphs from the population graph. In this paper, Metropolis-Hastings Random Walk (MHRW) and Random Walk with Jumps (RWwJ) sampling strategies are introduced, including the procedure of collecting nodes, the underlying mathematical theory, and corresponding estimators. We compared our methods and existing research outcomes and found that MHRW performs better when estimating degree distribution (61% less error than RWwJ) and graph order (0.69% less error than RWwJ), while RWwJ estimates follower and following ratio average and mutual relationship proportion in adjacent relationship with better results, with 13% less error and 6% less error than MHRW. We analyze the reasons for the outcomes and give possible future work directions.

## 1. Introduction

Online social network (OSNs) has drawn much attention in recent years due to the ever-increasing popularity of social media. Twitter, in particular, is one of the most important online social networks (OSNs) today. The large size dataset can be used to get information about marketing and people's behaviour, through some key characteristics.

Although these platforms can provide various and tens of information, some of them are too massive to store or analysis. Therefore, we need some sampling techniques to draw representative samples. In addition, appropriate model and estimators are also needed to estimate corresponding network parameters of interests. In this paper we focus on how to scale down the population graph. Therefore, we can get subgraphs that are small enough to be handled and preserve population graph properties.

In order to understand these 2 sampling strategies, the process of drawing samples and related estimators will be introduced in this paper. Their performances, estimated results and estimation errors would be presented precisely. Comparisons between sampling strategies are conducted after sampling experiments. Thus, we can figure out how different sampling strategies work in the same dataset while different graph parameters and analysis why these scenarios happen. The reminder of this paper will proceed as follows. In Section 2, related work will be reviewed. In Section 3 we introduce the methodology, including different sampling methods and estimation framework. In section 4, the details of simulation study are given, and the performances are evaluated, and Section 5 discusses, analyse and concludes.

## 2. Related Work

Random Walk (RW) is one of the most important and widely used sampling methods. From web pages network to social media networks, RW has a variety of usages in drawing samples from population graphs (Henzinger et al. 2000; Gkantsidis, Mihail, and Saberi 2004; Massoulié et al. 2006; Gjoka et al. 2010). Henzinger described how additional information obtained by the walk can be used to skew the sampling probability against pages with high PageRank. Gjoka developed a practical framework for obtaining a uniform sample of users in an online social network by crawling its social graph. Gkantsidis quantify the effectiveness of random walks for searching and construction of unstructured peer-to-peer (P2P) networks. Massoulié address the problem of counting the number of peers in a peer-to-peer system, and more generally of aggregating statistics of individual peers over the whole system.

However, RW is biased towards high degree nodes, which leads to many problems in estimation. Therefore, some modifications of random walk were proposed. One of the most useful method is Metropolis-Hastings Random Walk (MHRW), modifying the probabilities of next node selection in ordinary Random Walk (Hastings 1970). Using Metropolis Hastings method, uniform stationary distribution can be approximated by uanderlying Markov chain, thus the probabilities for visiting each node is uniformly distributed. With asymptotically equal probabilities, the sample mean is an asymptotically unbiased estimate of population mean.

Yet variants of RW have an obvious disadvantage. If the whole graph has two or more disconnected components, then MHRW can only reach equilibrium in a single component without jumps. Therefore, researchers proposed MHRW with jumps (Brin and Page 1998). Using jumps in the random walk based sampling strategies, the proposed distribution (for example, uniform distribution) can converge in the whole population graph.

## 3. Methodology

### *3.1 Problem formulation*

As Twitter friendship network can be modelled as a directed graph, let $G = (V, E)$ be the population graph, where $V$ is the nodes set of population graph, and $E$ is the edges set of population graph. Let $G_s = (V_s, E_s)$ be the subgraph of $G$, where $V_s$ is the nodes set of the subgraph, $E_s$ is the edge set of the subgraph. Edges between node $v_i$ and $v_j$ are denoted by $(ij)$, or $(ji)$. In directed graphs, edges $(ij)$ and $(ji)$ are different, since they have different directions. If both $(ij)$ and $(ji)$ exist in a directed graph, we would say that node $v_i$ and $v_j$ have a "mutual relationship". Graph order $N = |V|$ is the number of nodes (users) in $G$. Correspondingly, we assume that the node set of the subgraph $V_s$, has $n$ nodes in it, and the edge set is defined as $E_s = \{(ij) | \forall i, j \in V_s \text{ and } (ij) \in E\}$, representing interactions between nodes in the subgraph.

Matrix $a_{ij} \in A$ is a $N \times N$ matrix reviewing links between nodes. If edge $(ij)$ exists, $a_{ij} = 1$. Otherwise, $a_{ij} = 0$. Usually, $a_{ij} \equiv a_{ji}$ is invalid in the directed graph while it is true in an undirected graph. A motif is a property of a graph, denoted by $[M_k]$, where $M_k$ is relevant k-set of nodes.

4 targeted properties need to be estimated in this paper, including degree distribution, the total number of users, follower and following ratio, and the mutual relationship proportion in adjacent relationships.

### 3.1.1 Degree distribution

For any node $v_i$ in the graph, $(ij)$ is a directed edge from node $v_i$ to node $v_j$. In-degree of node $v_i$ is the total number of connections onto node $v_i$,

$$d^{in}(v_i) = \sum_j a_{ij} \qquad (1)$$

while the out-degree of node $v_i$ is the total number of connections coming from node $v_i$,

$$d^{out}(v_i) = \sum_j a_{ji} \qquad (2)$$

In-degree distribution is defined as

$$P_{deg}^{in}(k) = \text{fraction of nodes in the graph with in-degree } k \qquad (3)$$

Out-degree distribution is defined as

$$P_{deg}^{out}(k) = \text{fraction of nodes in the graph with out-degree } k \qquad (4)$$

### 3.3.2 Follower and following ratio

Having obtained in-degree and out-degree of nodes in graph, which represent the number of followers and number of followings correspondingly, we can calculate the ratio $R(v_i)$ of every node,

$$R_i = R(v_i) = \frac{d^{in}(k_i)}{d^{out}(k_i)} \qquad (5)$$

individual follower and following ratios can describe how user contributes to the platform. For the whole graph, the Follower and Following Ratio average is defined as:

$$\bar{R} = \frac{1}{n}\sum_{i=1}^{n} R_i \qquad (6)$$

follower and following ratio of one graph can describe how all users in the graph contribute to the platform averagely.

*3.1.3 Graph order*

In a graph model, graph order refers to the total number of nodes in the graph. In this paper, we denote the target graph order of the population graph as N, and the estimated order is denoted as $\hat{N}$.

*3.1.4 Mutual relationship proportion in adjacent relationships*

When both edges $(ij)$ and $(ji)$ exist in a directed graph, nodes $v_i$ and $v_j$ are friends, i.e., they have a mutual relationship.

Figure 1 Example Graph G

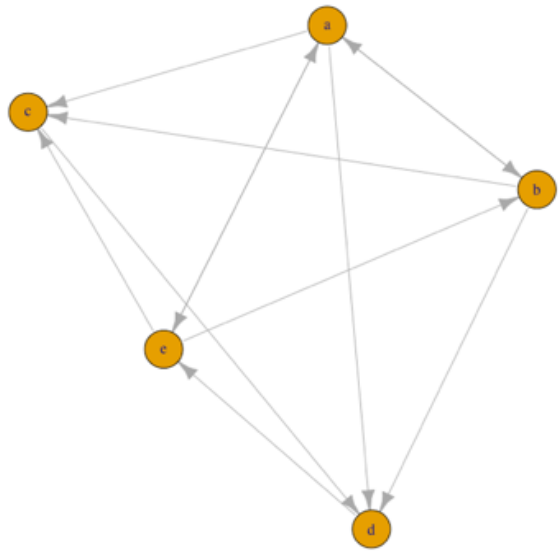

Mutual relationship is a second-order motif, which means that there are 2 nodes involved in the property of the graph. We would illustrate the concept of mutual relationship proportion in adjacent relationships in a simple graph. In Fig.1, the number of mutual relationships is 2.

There is a mutual relationship between node $a$ and node $b$. Another mutual relationship exists between node $a$ and node $b$. As for the proportion, it is 1/12 since there are 12 edges in total. Denote this proportion as $\sigma$, while the estimated value is $\hat{\sigma}$.

*3.2 Reviewing sampling methods*

In this paper, we mainly investigated two sampling methods, Metropolis-Hastings Random Work (MHRW) and Random Walk (RW). Previous research paid attention to them yet when we apply the link-tracing methods into massive social network datasets, we need to consider the social communities that are not connected or isolated nodes. Therefore, we also add the probability of jumping to prevent our walk from being stuck in a single community.

*3.2.1 Metropolis-Hastings Random Walk*

Firstly, we need to select an initial node, called "seed", from the population graph. Now, we have already got a starting node $s_0$, $Q$ is the transition matrix for a simple random walk with transition probabilities $q_{ij}$. Suppose that at step k, the state of the process is $X_k = v_i$, a provisional k+1th state is chosen among the rest of the nodes in population graph, by probabilities $q_{ij}$, $j = 1, 2, ..., N$ in the i*th* row of transition matrix $Q$. If the k+1th node is node $v_j$, compare the out-degrees of node $v_i$ and node $v_j$, which are denoted by $d_i$ and $d_j$. If $d_i > d_j$, node $v_j$ would be accepted as the k+1th state. Otherwise, we would take further steps that $p$ would be generated by a uniform distribution between 0 to 1, If $d_i/d_j$ is larger than $p$, node $v_j$ would be accepted as k+1th state. Otherwise, $v_j$ would be rejected and $v_i$ is accepted as k+1th state. Repeat these steps until the sample size b is reached. Using Metropolis-Hastings method (Hastings 1970), the transition matrix for the modified walk is denoted by $P_{ij}$, where

$$P_{ij} = q_{ij}\alpha_{ij}, for\ i \neq j \tag{7}$$

and

$$P_{ii} = 1 - \sum_{j \neq i} P_{ij} \tag{8}$$

where $\alpha_{ij} = \min\left(\frac{d_i}{d_j}, 1\right)$. Thus, we've got a uniform walk in a connected graph. However, when we want to get a uniform walk in a graph with many components, all the components are not connected in the graph, jumps are needed.

Let $d$ as the probability of walking while $1 - d$ is the probability of jumping. In the uniform walk with jumps, if there is an edge from node $v_i$ to $v_j$, the acceptance probability can be modified as

$$\alpha_{ij} = \min\left(\frac{\frac{1-d}{N} + \frac{d}{d_j}}{\frac{1-d}{N} + \frac{d}{d_i}}, 1\right) \tag{9}$$

In addition, if there is no edge from node $v_i$ to $v_j$, the acceptance probability is $\alpha_{ij} = 1 - d$. Other acceptance probabilities have $\alpha_{ij} = 1$.

When Uniform Walk draws a candidate node $v_j$, one of the neighbours of node $v_i$, with probability $1/d_i$, the Uniform Walk algorithm also accepts $v_j$ into the sample with probability $\min(d_i/d_j, 1)$, while rejects with probability $1 - \min(d_i/d_j, 1)$. The stationary distribution is the uniform distribution

$$\pi = \left(\frac{1}{N}, \frac{1}{N}, \dots, \frac{1}{N}\right) \tag{10}$$

When using MHRW, the sample mean can be used as an unbiased estimator of the mean of all the nodes. The Metropolis-Hastings Random Walk algorithm details are given in Algorithm 1 in the appendix.

*3.2.2 Random Walk*

The random Walk method has been used in many graph sampling procedures, while walking can only reach nodes with positive in-degree in directed graphs. If the starting node has no edges with other nodes, the equilibrium can be reached only on one component of the whole graph. To prevent this situation, we set a probability of jumping, selecting nodes randomly if necessary.

when our current state is node $v_i$, we have an out-degree$+\alpha$ choice(s) to choose a node (including the virtual node) to reach. Thus, the probability of choosing the neighbour $v_j$ is $w_{ij}/(d_i^{out} + \alpha)$, where $w_{ij}$ is the edge weight of $(ij)$ and $d_i^{out}$ is out-degree of node $v_i$. When choosing an existing node, $w_{ij} = 1$, while $w_{ij} = \alpha$ when walking to the virtual node. The former case is called the walking case, in which one of the neighbours of node $v_i$ is collected with probability $1/(d_i^{out} + \alpha)$. The latter case is jumping. Apart from choosing the jumper, a node in the graph should be selected uniformly, in which probability is $1/N$. Thus, the probability of jumping into a random node is $\frac{\alpha}{(d_i^{out}+\alpha)} \cdot (\frac{1}{N})$. Under the jumping condition, the probability of choosing a node connected with $v_i$ is $\frac{\frac{\alpha}{N}+1}{(d_i^{out}+\alpha)}$, for a node that is not a neighbour of $v_i$, the probability of being k+1th state is $\frac{\alpha/N}{(d_i^{out}+\alpha)}$. Thus, the transition matrix of underlying Markov chain is that

$$q_{ij} = \begin{cases} \frac{\frac{\alpha}{N}+1}{(d_i^{out}+\alpha)}, & if\ (ij) \in E \\ \frac{\frac{\alpha}{N}}{(d_i^{out}+\alpha)}, & if\ (ij) \notin E \end{cases} \quad (11)$$

Since the transition matrix satisfy the property that the sum of all elements in one row equals 1, i.e.

$$\sum_{j=1}^{N} q_{ij} = 1 \quad (12)$$

$q_{ij}$ also satisfies the property. Here we divide all nodes in the original graph into two node sets, for any node $v_i$,

$$S_1 = \{v_j | (ij) \in E\} \tag{13}$$

and

$$S_2 = \{v_j | (ij) \notin E\} \tag{14}$$

Where edges $(ij)$ are all directed edge from node $v_i$ to node $v_j$.

$$
\begin{aligned}
\sum_{j=1}^{N} q_{ij} &= \sum_{(ij)\in E} q_{ij} + \sum_{(ij)\notin E} q_{ij} \\
&= \sum_{(ij)\in E} \frac{\frac{\alpha}{N}+1}{d_i^{out}+\alpha} + \sum_{(ij)\notin E} \frac{\frac{\alpha}{N}}{d_i^{out}+\alpha} \\
&= |S_1|\frac{\frac{\alpha}{N}+1}{d_i^{out}+\alpha} + |S_2|\frac{\frac{\alpha}{N}}{d_i^{out}+\alpha} \\
&= \frac{\frac{\alpha}{N}(|S_1|+|S_2|)+|S_1|}{d_i^{out}+\alpha} \\
&= \frac{\alpha+|S_1|}{d_i^{out}+\alpha} \\
&= 1
\end{aligned}
\tag{15}
$$

Where sets $|S_1|$ and $|S_2|$ are the order of set $S_1$ and $S_2$. Note that $|S_1| = d_i^{out}$, the out-degree of node $v_i$ is the number of nodes that linked with a directed edge comes from node $v_i$. Thus, we can get 1 in the last step. Also, the node sets $S_1$ and $S_2$ are disjointed because one node can only be the neighbour of node $v_i$ or not. Thus, those two node sets are exhaustive and mutually disjointed. $|S_1| + |S_2| = N$ is valid in the proof.

According to previous research (Zhao et al. 2019), the underlying Markov chain of RW satisfies the stationary condition, so that RWwJ converges to a stationary distribution. However, we cannot calculate the exact values of inclusion probabilities. and the probability of $v_i$ being chosen is proportional to in-degree $d_i^{in} + \alpha$ in the stationary distribution.

So, the generalized ratio estimator of nodal parameters can be given as

$$\hat{\theta} = \frac{\sum_{v_i \in V_s} f(v_i)/(d_i+\alpha)}{Z^{RWwJ}} \tag{16}$$

where $f(v_i)$ is defined as a nodal parameter and $Z^{RWwJ} = \sum_{v_i \in V_s} 1/(d_i + \alpha)$. In this paper, all the parameters can be defined as nodal parameters. Since the limiting distribution is used as sampling distribution from which we draw samples, the estimator is asymptotically unbiased.

Proof: Suppose function $f: V \rightarrow \mathbb{R}$, $\theta \triangleq \frac{1}{N}\sum_{i=1}^{N} f(v_i)$, $\theta$ is the mean of a population graph property. According to large number law, the sample mean is equal to the population mean when sample size grows up, i.e.

$$\lim_{n \rightarrow N} \frac{1}{n} \sum_{i=1}^{n} f(x_i) = E_\pi[f(x_i)] = \sum_{i=1}^{N} [\pi_i f(x_i)] \tag{17}$$

where $E_\pi[f]$ is the expectation of population mean and $\pi$ is the stationary distribution of the population graph. Similarly, function g on random variable also has property

$$\lim_{n \rightarrow N} \sum_{i=1}^{n} g(x_i) = E_\pi[g(x_i)] = \sum_{i=1}^{N} [\pi_i g(x_i)] \tag{18}$$

then,

$$\lim_{n \rightarrow N} \frac{\sum_{i=1}^{n} f(x_i)}{\sum_{i=1}^{n} g(x_i)} = \frac{\lim_{n \rightarrow N} \sum_{i=1}^{n} f(x_i)}{\lim_{n \rightarrow N} \sum_{i=1}^{n} g(x_i)} = \frac{E_\pi[f]}{E_\pi[g]} \tag{19}$$

therefore, when replace $f(x_i)$ and $g(x_i)$ with $f(v_i)/(d_i^{in} + \alpha)$ and $1/(d_i^{in} + \alpha)$, then $\frac{\sum_{v_i \in V_s} y(v_i)/(d_i^{in}+\alpha)}{\sum_{v_i \in V_s} 1/(d_i^{in}+\alpha)} \rightarrow \theta$ as $n \rightarrow N$. Thus, generalized ratio estimator for the degree distribution and follower and following ratio in RWwJ is an asymptotically unbiased estimator. Details of the Random Walk with Jumps algorithm are in the appendices.

### 3.3 Estimators

Since Uniform Walk is implemented, the estimated sample mean would be an unbiased estimator of the population mean.

$$\hat{\theta} = \frac{1}{n} \sum_{v_i \in V_s} f(v_i) \tag{20}$$

The assumptions of estimating each parameter are independent. When the degree distribution, follower and following ratio average, and mutual relationship proportion are estimated, the total number of nodes in the population graph is known, while the population order is unknown when the population graph order is estimated.

For degree distribution, let function $f(v_i) = \mathbf{1}_{\{d_i=k\}}$. This is an indicator function. When the degree is k, the function value is 1. Otherwise, the function value equals zero. The estimated proportion of nodes with k is given by

$$\hat{\theta}_k = \frac{1}{n} \sum_{i=1}^{n} \mathbf{1}_{\{d_i=k\}} \tag{21}$$

both unbiased for in-degree and out-degree distribution.

For average follower and following ratio, let function $f(v_i) = \mathbf{1}_{\{R_i=k\}}$, where $R_i$ is the follower and following ratio of node $v_i$. Therefore, the estimated proportion of node with ratio k is given by

$$\hat{\omega}_k = \frac{1}{n} \sum_{i=1}^{n} \mathbf{1}_{\{R_i=k\}} \tag{22}$$

For graph order, the capture and recapture method is used (Kurant, Butts, and Markopoulou 2012). In the Uniform Walk with Jumps method, we independently collect two uniform samples $S_1^{\text{uniq}}$ and $S_2^{\text{uniq}}$ without replacement. The population graph order can then be estimated by

$$\hat{N} = \frac{|S_1^{\text{uniq}}| \cdot |S_2^{\text{uniq}}|}{|S_1^{\text{uniq}} \cap S_2^{\text{uniq}}|} \tag{23}$$

Considering the Metropolis-Hastings Random Walk process might have repeated nodes, we can apply this estimator to a uniform sample $S$ by randomly splitting $S$ into two equal-sized subsamples $S_1$ and $S_2$, and then discarding the repetitions within each of them to obtain $S_2^{\text{uniq}}$ and $S_1^{\text{uniq}}$.

For mutual relationship proportion in adjacent relations,

$$\hat{\sigma} = \pi_i q_{ij} + \pi_j q_{ji} \qquad (24)$$

where $\pi_i$ is the inclusion probability. For a uniform walk, the inclusion probability is asymptotically equivalent to $1/N$; $q_{ij}$ is the transition probability of node $v_i$ to $v_j$. Therefore, MHRW based estimator can be rewritten as

$$\hat{\sigma} = \frac{1}{N}(q_{ij} + q_{ji}) \qquad (25)$$

Since MHRW is targeted at uniform distribution, in which all the inclusion probabilities of nodes are $\frac{1}{N}$. In the experiments, an estimator given by (25) will be used.

As given in the previous context, the estimator is given by the formulation mentioned above. For degree distribution, let function $f(v_i) = \mathbf{1}_{\{d=k\}}$. The estimator of the proportion of nodes that have k-degree is given by

$$\hat{\theta}_k = \frac{\frac{\sum_{v_i \in V_s} f(v_i)}{d_i^{in} + \alpha}}{Z^{RWwJ}} = \frac{\frac{\sum_{v_i \in V_s} \mathbf{1}_{\{d=k\}}}{d_i^{in} + \alpha}}{Z^{RWwJ}} \qquad (26)$$

both unbiased for in-degree and out-degree distribution.

For average follower and following ratio, let function $f(v_i) = \mathbf{1}_{\{R=k\}}$, where $R_i$ is a follower and following ratio of node $v_i$. Therefore, the estimated proportion of node with ratio k is given by

$$\hat{\omega}_k = \frac{\frac{\sum_{v_i \in V_s} f(v_i)}{d_i^{in} + \alpha}}{Z^{RWwJ}} = \frac{\frac{\sum_{v_i \in V_s} \mathbf{1}_{\{R_i=k\}}}{d_i^{in} + \alpha}}{Z^{RWwJ}} \qquad (27)$$

For graph order, same as the Uniform Random Walk, the capture and recapture method is used. In Random Walk with Jumps method, we independently collect two samples $S_1$ and $S_2$ without replacement, where one of the samples should be a uniform sample(Kurant, Butts, and Markopoulou 2012).

Since Random Walk with jumps cannot draw a uniform sample from the designed sampling procedure, there must be one uniform sample, which can be provided with MHRW mentioned above. let $S_1$ be the uniform sample and $S_2$ be the sample drawn in any arbitrary way. We use RWwJ. It is obvious that $S_2$ can be repeated. After getting those two sets, we can calculate the number of cross-collisions $n^{xcol}$ between $S_1$ and $S_2$.

$$n^{xcol} = \sum_{s_1 \in S_1} \sum_{s_2 \in S_2} \mathbf{1}_{\{s_1 = s_2\}} \tag{28}$$

since every node in $S_1$ is selected uniformly from all $N$ nodes, the probability that $s_1 \in S_1$ collides with a given $s_2 \in S_2$ is

$$\Pr(s_1 = s_2) = \frac{1}{N} \tag{29}$$

so, the expected number of collisions is

$$E[n^{xcol}] = \sum_{s_1 \in S_1, s_2 \in S_2} Pr(s_1 = s_2) = \frac{|S_1||S_2|}{N} \tag{30}$$

by replacing $E[n^{xcol}]$ with the value $n^{xcol}$ measured in reality, we obtain the following estimator

$$\widehat{N} = \frac{|S_1||S_2|}{n^{xcol}} \tag{31}$$

This resembles the capture and recapture method, except that here only one set $S_1$ is required to be uniform, and the other one set $S_2$ is arbitrary. Moreover, we allow for repetitions.

When estimating Mutual Relationship Proportion in Adjacent Relationship under RWwJ, there is something different from the Uniform Walk, RWwJ has no inclusion probabilities. However, as we have known the population order $N$, then we can standardize the probabilities using $1/(d_i^{in} + \alpha)$. Therefore, the weights can be written as

$$w_i = \frac{\frac{1}{d_i^{in} + \alpha}}{Z^{RWwJ}} \tag{32}$$

Same as the mutual relationship proportion estimated in the uniform walk, the estimation based on RWwJ is $\hat{\theta} = w_i q_{ij} + w_j q_{ji}$.

Although we can get the standardized probability of nodes in the population graph, the standardized probability is not the real inclusion probability. That's the reason why total (asymptotically) unbiased estimators cannot be derived using Random Walk.

## 4. Simulation

### *4.1 Data*

The dataset is downloaded from a website from the laboratory (SNAP: Network datasets: Higgs Twitter Dataset, 2022), which is a dataset used for academic research. The dataset has been built after monitoring the spreading processes on Twitter before, during, and after the announcement of the discovery of a new particle with the features of the elusive Higgs boson on 4th July 2012. The messages posted on Twitter about this discovery between 1st and 7th July 2012 are considered.

### *4.2 Implementation details*

To compare the algorithms' performances, the cost is controlled in our experiments, which is 62072. This is based on the previous research (Leskovec and Faloutsos 2006). They conducted experiments and concluded that the best sample size is 15% of the original graph through conducting experiments and summarizing the results.

Since we have 456626 in the population graph, 15% of 456626 is 62072. For these 2 sampling strategies based on a random walk, although total cost b = 62072, i.e., we repeat the collection process 62072 times. The subgraphs sampled by MHRW and RWwJ algorithms have 54339 nodes and 56973 nodes correspondingly, which are less than the cost we set.

The reason why we only collected node sets that are smaller than cost is that both MHRW and RWwJ are RW-based sampling methods, some of the nodes may be collected more than once.

Particularly, MHRW always collects fewer nodes than RWwJ because MHRW has an acceptance probability, which might reject the candidate nodes and stay in the current node until it chooses to jump or have an acceptable candidate node in walking procedure, yet RWwJ does not have such a drawback. Therefore, controlling cost in the sampling method is an important and fair condition to compare different sampling strategies.

The subgraph consists of the node set and edge set. For RW-based strategies, edge sets are collected by walking procedure when sampling methods run on population graph. if current node $v_i$ selects jumping to a random node $v_j$ uniformly, then there is no edge collected, even if there is a directed edge $(ij)$.

### *4.3 Results*

#### *4.3.1 Degree Distribution*

The degree distribution provides the most basic summary of the connectivity of nodes in graphs. As we can see in the results (Fig. 2 and Fig. 3), one common aspect is that nodes with high degrees, both in-degree and out-degree, only have a small fraction of the whole users. Most of the nodes, which represent users in the social media platform, only have very few links among them. These empirical distributions are consistent with most studies in degree distribution.

Figure 2 Cumulative Frequency of Estimated In-Degree Compared with Population graph

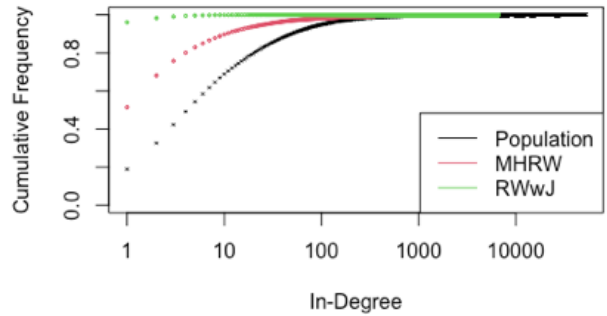

Figure 3 Cumulative Frequency of Estimated Out-Degree Compared with Population graph

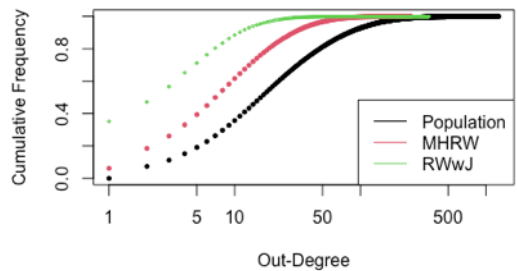

tables of D-statistics on the candidate sampling strategies. From the definition of D-statistics, the smaller D-statistics are, the better candidate sampling strategies are. From the results shown in the table, MHRW has smaller D-statistics, in both estimating in and out-degree distribution. MHRW performs better than RWwJ. However, for in-degree distribution, the two sampling strategies' performances difference is relatively small, MHRW has 0.9052 D-statistics, which means that the maximal estimated degree distribution error is 0.9052, while RWwJ has 0.98116 maximum error. For out-degree estimation, MHRW did better than RWwJ, 0.28571 against 0.9998. It is quite a significant outperformance.

Table 1 D-statistics between MHRW Sample and Population Graph and Total Variation Distance between RWwJ Sample and Population Graph in Degree Distributions

| **Distribution** | MHRW | RWwJ |
| --- | --- | --- |
| In-degree | 0.9052 | 0.98116 |
| Out-degree | 0.28571 | 0.99998 |

Similarly, the second evaluation of distribution properties is KL divergence, of which evaluation results are presented below. It values the sum of logarithms of differences between the population in/out-degree distributions and sample in/out-degree distributions. In KL divergence, the performance differences between MHRW and RWwJ are not so significant as it is valued by D-statistics. The reason why is that the evaluation way of KL divergence uses a logarithm function, which is smaller than the linear function in D-statistics, i.e., absolute value function. Furthermore, out-degree estimation has less error, the same as the D-statistics outcome.

Table 2 KL Deviance of between MHRW Sample and Population Graph and Total Variation Distance between RWwJ Sample and Population Graph in Degree Distributions

| **Distribution** | MHRW | RWwJ |
|---|---|---|
| In-degree | 0.01244723 | 0.01248951 |
| Out-degree | 0.0001300872 | 9.805315e-05 |

the CDF plots show that MHRW can estimate degree distributions better, which is consistent with the conclusions drawn from the above statistics outcomes.

Real estimated values (fraction of node with degree $k$) are bigger than the original values in the population graph. The reason why estimated values are larger is that the Random Walk algorithm tends to collect higher degree nodes (Zhao 2012). As nodes with a higher degree are more likely to be collected, the sample is also biased towards higher-degree nodes. Although weights are assigned in the sample, it is still not enough to correct the bias. Using a biased sample will lead to biased results.

As for MHRW, it corrects bias when collecting the sample, changing the transition probabilities by setting acceptance probabilities. But then it also has convergence efficiency problems. Previous research conducted by Gjoka (2011, Fig.3) has compared the efficiency of convergence between Random Walk and the Metropolis-Hastings Random Walk. The conclusion is that Metropolis-Hastings Random Walk has worse convergence efficiency than

Random Walk. Poor convergence efficiency means that there are a large proportion of nodes and edges sampled by MHRW is not uniformly distributed although the Metropolis-Hastings method has promoted the acceptance probability. This could be the main reason why the results have a bias in estimating degree distribution.

*4.3.2 Follower and following Ratio Average*

Following means how many accounts the user follows on Twitter, while the follower represents how many accounts follow the user. They are denoted as out-degree and in-degree of nodes in the graph. So the follower and the following ratio is an important measurement that speaks a lot about the behaviour of a Twitter user and is also a measure of how well the users contribute to Twitter (Salehi 2011). Here the average of all users' followers and following ratios can reflect how well the users included in the graph contribute to the platform.

The experiment result shows that the MHRW method estimated the value as 1.83 while RWwJ estimated this parameter as 2.5. In the population graph, this parameter is 5.1. Therefore, we can evaluate the errors. The relative root mean square error of the estimated values are 64% and 51%, respectively. The outcomes of those two sampling strategies underestimated this property, however, RWwJ is relatively closer.

This outcome is consistent with previous research (Salehi 2011), MHRW does not perform well in the follower and following ratio distribution. After estimating the ratio distribution by MHRW, the outcome is consistent with the previous research. That is the reason why average ratio estimation has so much error in the experiment.

*4.3.3 Graph Order*

Using the capture and recapture method mentioned above, the estimated value of graph order by Uniform Walk is 433055, while Random Walk estimates it as 427771. Compared with the known population graph total 456626, the MHRW strategy underestimated the graph order, of which the relative root mean square error is 0.0516. RWwJ has a slightly larger error, which is 0.063.

Table 4 Estimation of Graph Order and Relative Root Mean Square Error

|  | Population graph | MHRW | RWwJ |
|---|---|---|---|
| Order/Estimated Order | 456626 | 433055 | 427771 |
| Relative Root Mean Square Error | / | 0.0516 | 0.063 |

These two results are calculated according to corresponding estimators. From the theory part, we can see that at least one capture should be uniformly distributed.

Graph order is studied in many prior works. Different from non-graphic structured data, biological research on population total using capture-recapture method focus on the assumptions that the population is approximately closed over the secondary sampling periods within a primary period (Pollock 1982). If there is a significant increase in the population, then the capture and recapture method will have a very large error compared with the real population total. Only if the population total is constant, the two estimation procedures are possible. The most common sampling method in estimating creature population total is the simple random sample (SRS) strategy in each capture procedure. SRS strategy guarantees the estimator is valid and effective.

In estimating a network structured data population total (population graph order), if we use a uniform node sampling method, the results are the same as the ordinary data. However, when we consider the links between objective/ individuals in the graphs, there is dependence between node and node, which is not a simple random sample for nodes. But MHRW overcome this disadvantage in graph sampling to some extent. In MHRW methods used in the dissertation, the target distribution is uniform, followed by the nodes' frequencies when the Markov chain becomes stationary.

*4.3.4 Mutual Relationship Proportion in Adjacent Relationship*

To calculate this proportion, the number of mutual relationships and all adjacent relationships should be counted. In the population graph, the proportion is 0.31, which means that there are 31% of users are friends among all the adjacent relationships.

The estimated proportion by MHRW is 0.16, while the RWwJ estimate of proportion is 0.18. Both results indicate that the proportion is underestimated. For evaluation, the errors are calculated to be 48% and 42% correspondingly.

Table 5 Mutual Relationship Proportion in Adjacent Relationship

| Mutual relationship proportion | Population Graph | MHRW | RWwJ |
|---|---|---|---|
| Estimated Value | 0.31 | 0.16 | 0.18 |
| Relative Root Mean Square Error | / | 48% | 42% |

From the estimated results we can see that the probability of mutual relationship is lower than that in the population graph. The reason why we would get a lower proportion is that we add jumps into the sampling methods to overcome the disconnected components of the whole population graph. When the algorithm is in the kth step of collecting nodes and edges, the current status is node $v_i$, even if the out-degree of $v_j$ is not equal to zero, i.e., $v_i$ is not an isolated node, there is still a probability of jumping in the *k+1*th step by choosing a node $v_j$ uniformly. If there is an edge between $v_i$ and $v_j$, the edge would not be included in the sample edge set. Therefore, if the node is selected by jumping, the edge can be excluded from the edge set. This will lead to an underestimated number of the subgraph consisting of sampled edges and nodes, which leads to the underestimation of the target proportion.

Although jumps can overcome the problem of disconnected components of the population graph, they also may lead to the underestimation of edges in the subgraphs. However, the algorithm can only work and converge to stationary distributions in the component where the starting node lies without jumps. This may lead to a small and repeated sample size, which is a waste of cost in the sampling procedure. A balance of the sample size and estimation should be proposed when measuring the graph parameters related to the edge.

## 5. Conclusion

The paper aims to use graph sampling methods to draw subgraphs and use the sample (the subgraph) to estimate target characteristics of the population graph and compare their performances. There are 4 characteristics of interest, degree distributions, follower and following ratio average, graph order, and the mutual relationship proportion in adjacent relationships. These are important characteristics that can describe social networks. Since previous studies show that some RW-based sampling methods perform well in social network data, Metropolis-Hastings Random Walk and Random Walk with Jumps are chosen in the paper. We use these two sampling methods to draw samples from the real-world social network, which is conceptualized as a graph in our study.

The experimental results of degree distributions in the dissertation, shown in Fig. 9, indicate that a significant fraction of nodes in the graph have very low in-degree and out-degree, which means most Twitter users have a low number of followers/ followings. This is a very common property among social networks (Rejaie et al. 2010). The estimated results of the degree distribution, both in-degree and out-degree, show that MHRW performs better than the RWwJ, this is consistent with the conclusion drawn by (Gjoka et al. 2010). In their paper, MHRW did better than RW when estimating degree distributions.

The estimation of follower and following ratio average shows that RWwJ outperforms MHRW in estimating the follower and following ratio average. For MHRW, the error is 64% while the error of RWwJ estimation is 51%. Although RWwJ outperforms MHRW, its estimation also has a large variance with the average population follower and following ratio. To get an accurate estimation, more modifications should be proposed on the sampling methods or corresponding estimators.

Network size estimation shows that the capture-recapture method can be used in estimating the social network total users, with relatively small errors. With different distributed samples, different ways of tackling samples are used in our dissertation. Experimental results in previous work also show verified this conclusion(Kurant, Butts, and Markopoulou 2012; Lu et al. 2021). The mutual relationship proportion in adjacent relationships estimation uses a different way to estimate. When estimating the first 3 graph parameters, we collect the sample, construct estimators, and use the subgraphs to get the estimates. But for a mutual relationship, the way of estimating the parameter is using the inclusion probabilities to multiply the transition probability as long as nodes are collected. In the experimental results, RWwJ shows better performance than MHRW.

Even if the MH algorithm has been widely used since it can draw an unbiased sample. This "ready to use" portraits attract much attention to the algorithm, but it is not the only solution to estimation. Much related research also has shown that and use other Random Walk variants to do graph sampling (Al Hasan and Zaki 2009; Gjoka et al. 2011; Ahmed, Neville, and Kompella 2012).